\documentclass{article}

\usepackage{microtype}
\usepackage{graphicx}
\usepackage{subfigure}
\usepackage{booktabs} % for professional tables
\usepackage{float}
\usepackage{multirow}
\usepackage[table,xcdraw]{xcolor}

\usepackage{hyperref}

\usepackage{physics}
\usepackage{amsfonts}

\usepackage[accepted]{icml2021}

\icmltitlerunning{Orbital Mixer}

\begin{document}

\twocolumn[
\icmltitle{Orbital Mixer: Using Atomic Orbital Features for Basis Dependent Prediction of Molecular Wavefunctions}



%\icmlsetsymbol{work}{*}
\icmlsetsymbol{internship}{*}

\begin{icmlauthorlist}
\icmlauthor{Kirill Shmilovich}{pme,internship}
\icmlauthor{Devin Willmott}{bcai}
\icmlauthor{Ivan Batalov}{bcai}
\icmlauthor{Mordechai Kornbluth}{rtc}
\icmlauthor{Jonathan Mailoa}{tencent}
\icmlauthor{J. Zico Kolter}{bcai,cmu}
\end{icmlauthorlist}

\icmlaffiliation{pme}{Pritzker School of Molecular Engineering, University of Chicago, Chicago, IL, United States}
\icmlaffiliation{bcai}{Bosch Center for Artificial Intelligence, Pittsburgh, PA, United States}
\icmlaffiliation{rtc}{Bosch Research and Technology Center, Cambridge, MA, United States}
\icmlaffiliation{tencent}{Tencent Quantum Laboratory, Shenzhen, Guangdong, China}
\icmlaffiliation{cmu}{Carnegie Mellon University, Pittsburgh, PA, United States}

\icmlcorrespondingauthor{Kirill Shmilovich}{kirills@uchicago.edu}

\icmlkeywords{}

\vskip 0.3in
]



\printAffiliationsAndNotice{\internship} % otherwise use the standard text.

\begin{abstract}
Leveraging \textit{ab initio} data at scale has enabled the development of machine learning models capable of extremely accurate and fast molecular property prediction. A central paradigm of many previous works focuses on generating predictions for only a fixed set of properties. Recent lines of research instead aim to explicitly learn the electronic structure via molecular wavefunctions from which other quantum chemical properties can directly be derived. While previous methods generate predictions as a function of only the atomic configuration, in this work we present an alternate approach that directly purposes basis dependent information to predict molecular electronic structure. The backbone of our model, Orbital Mixer, uses MLP Mixer layers within a simple, intuitive, and scalable architecture and achieves competitive Hamiltonian and molecular orbital energy and coefficient prediction accuracies compared to the state-of-the-art.    
\end{abstract}

\section{Introduction}
\label{sec:intro}

An explosion of interest has surrounded applying machine learning (ML) methods to quantum chemistry with a plethora of interesting application areas such as learning interatomic potentials~\cite{behler2007generalized,unke2021machine,bartok2010gaussian,smith2017ani,chmiela2017machine,chmiela2018towards, schutt2018schnet, unke2019physnet, unke2021spookynet, batzner2021se, klicpera2020directional, liu2021spherical, schutt2021equivariant}, constructing density functionals~\cite{snyder2012finding,brockherde2017bypassing,ryczko2019deep,kalita2021learning,li2021kohn}, predicting spectroscopic properties~\cite{gastegger2017machine,westermayr2020deep}, optoelectronic properties~\cite{lee2021transfer,mazouin2021selected, lu2020deep, gladkikh2020machine}, activation energies~\cite{lewis2021machine, grambow2020deep}, and a variety of physical properties throughout chemical compound space~\cite{montavon2013machine,de2016comparing,von2020exploring,keith2021combining,liu2021transferable,tielker2021quantum,Bratholm2021}. Quantum chemistry workflows can obtain such chemical and physical information by modelling the electronic Schrodinger equation in a chosen basis set of localized atomic orbitals that is then used to derive the ground-state molecular wavefunction. Using ML, we can rather directly predict the molecular electronic structure which then provides access to a plethora of these derived properties without needing to train specialized models for each property of interest. Previous works of Schütt et al.~\cite{schutt2019unifying} (SchNOrb) and most recently Unke et al.~\cite{unke2021se} (PhiSNet) present deep learning architectures for predicting molecular wavefunctions and electronic densities by purposing only information of the atomic coordinates and molecular composition. Though inputs to these models rely only on the raw features of the molecule, they are trained on molecular wavefunctions from real quantum chemistry calculations, which necessarily associates the model's predictions with a prescribed basis.

%In Schütt et al.~\cite{schutt2019unifying} the authors present a deep learning framework, dubbed SchNOrb, that performs this task as a function of only the atomic coordinates and molecular composition, without purposing any information of the atomic orbital basis functions. \noter{Seems like this paragraph would benefit from some concluding sentense here -- not sure exactly what it should be without just repeating the start of the next paragraph?}

Here, we present an alternate approach for predicting molecular orbitals that explicitly supplies basis set-specific information as input to a deep learning architecture that models interactions of atomic orbitals representations. Compared to only atomic coordinates and molecular composition, a complete basis set can provide a much higher dimensional and information rich representation of a molecular configuration~\cite{qiao2020orbnet}. Our model, called Orbital Mixer, purposes characteristics of the atomic orbitals jointly with their spatial overlap to predict the orbital coefficients that define the molecular electron density. The backbone of the Orbital Mixer architecture uses MLP mixer~\cite{mlpmixer} layers to efficiently model interactions between atomic orbital representations, and to ultimately predict the electronic Hamiltonian $\mathbf{F}$ for a molecular configuration, which can be diagonalized to obtain molecular orbitals. Directly operating on atomic orbital representations provides a strong inductive bias for Orbital Mixer when predicting the Hamiltonian, $\mathbf{F}$ which is represented in the same atomic orbital basis.       

We evaluate Orbital Mixer against SchNOrb and PhiSNet on three separate molecular configuration datasets. Similar to SchNOrb, but unlike PhiSNet, Orbital Mixer is not explicitly covariant with respect to rigid molecular rotations but rather is trained using data augmentation to learn this equivariance, in line with previous interatomic-potential work, although explicitly-covariant schemes can also be developed \cite{Mailoa2019,Park2021}. We report improved accuracy and data and parameter efficiency when predicting the electronic Hamiltonian $\mathbf{F}$, molecular orbital coefficients and energies compared to SchOrb while also performing competitively compared to PhiSNet. We demonstrate how integrating Orbital Mixer into quantum chemistry workflows using the predicted Hamiltonian $\mathbf{F}$ as an initial guess to DFT calculations achieves improved convergence speeds compared to default methods. Lastly, directly predicting the electronic structure gives us access to a variety of derivable physical chemical properties without needing to train separate ML models for each property of interest. We show excellent agreement between Orbital Mixer predicted and reference calculations for HOMO-LUMO gap energies and electronic dipole moments. The Orbital Mixer architecture benefits from simple and intuitive construction while leveraging strong inductive biases operating directly on atomic orbital representations to achieve competitive prediction accuracies.

%Our model is evaluated against baseline SchNOrb performance for three separate molecule configuration datasets. We report across the board improved prediction accuracy for the electronic Hamiltonian $\mathbf{F}$, molecular orbital coefficients and energies when training on using $\sim$10$\times$ fewer samples and containing $\sim$9$\times$ fewer model parameters than SchNOrb. We demonstrate how integrating our model into quantum chemistry workflows using our predicted Hamiltonian $\mathbf{F}$ as an initial guess to DFT calculations achieves improved convergence speeds compared to default methods. Lastly, directly predicting the electronic structure gives us access to a variety of derivable physical chemical properties without needing to train separate ML models for each property of interest. We show excellent agreement between our model predicted and reference calculations for HOMO-LUMO gap energies and electronic dipole moments, once again outperforming SchNOrb with fewer training samples and model parameters. The scalability and accuracy of our approach presents an important step towards tighter integration of ML within quantum chemistry workflows.     

\section{Background and related work}

\subsection{Deep Learning for Orbital Prediction}

Machine learning for molecular inference has experienced impressive success in recent years, showcasing spectacular predictive accuracy enabled by large quantities of \textit{ab initio} data, the incorporation of prior physical and chemical knowledge, and invariant and/or equivariant architectures~\cite{keith2021combining}. A common paradigm of these works interprets molecules as connected graphs and uses message passing to model interactions as a function of single-particle contributions. A variety of increasingly complex graph convolutional operations have been proposed for these purposes, such as the pioneering work of SchNet~\cite{schutt2018schnet} introducing continuous filter convolutions, the message passing designed based on physical principles and attention of PhysNet~\cite{unke2019physnet}, the explicitly covariant network operations of Cormorant~\cite{anderson2019cormorant}, Tensor field networks~\cite{thomas2018tensor} and NequIP~\cite{batzner2021se}, among others~\cite{schutt2021equivariant,klicpera2020directional,schutt2017quantum,klicpera2021gemnet,liu2021spherical,haghighatlari2021newtonnet,unke2021spookynet}. While these works have demonstrated excellent expressively and accuracy for molecular property prediction, each network is trained to predict only a predetermined set of scalar, vector, or sometimes tensor quantities. This fundamental design of these networks therefore requires training separate bespoke models for each molecular property of interest. 

In contrast, a recent line of research strives instead to ascertain molecular wavefunctions by predicting the Hamiltonian matrix that satisfies the electronic Schrodinger equation from which physical and chemical properties can be derived. The short history of these methods begins with Hedge and Bowen~\cite{hegde2017machine}, where they predict the Hamiltonian for two simple Copper and Carbon (diamond) systems using kernel ridge regression. Schütt et al.~\cite{schutt2019unifying} then proposed a deep learning architecture called SchNet for Orbital (SchNOrb) that uses the SchNet architecture and pair-wise features to predict the Hamiltonian block-wise, establishing baselines for molecule configurations from the popular MD17 dataset~\cite{chmiela2017machine}. Follow-up work from Gastegger et al.~\cite{gastegger2020deep} reports improved accuracy on select molecules by applying SchNOrb trained on a minimal basis set representation of molecular wavefunctions. More recently, Unke et al.~\cite{unke2021se} propose PhiSNet, which draws upon insights of SE(3)-equivariant models to maintain that Hamiltonian predictions remain explicitly covariant with respect to rigid rotations or translations while also reporting significantly improved prediction accuracies. Notably, Nigam et al.~\cite{nigam2021equivariant} devise similarly equivariant Hamiltonian representations for uses in other applications such as kernel machines. 

\subsection{Modeling Atomic Interactions}

Architectural choices around modeling interactions between atoms and other molecular are of central interest in the design of neural networks for molecular inference. The aforementioned common choices of graph neural network structure or convolution/mixing operation dependent on (pairwise) atomic distances have the benefit of incorporating geometric information into the structure of the network and prioritizing local interactions. These approaches come at the cost of additional hyperparameters or increased network complexity, and frequently requires imposing a distance cutoff that prevents the network from directly modeling long-range atomic interactions~\cite{unke2019physnet}.

In lieu of a graph neural network architecture, we draw on approaches for mixing spatial information from other deep learning domains, namely those from computer vision, where vision transformers (ViTs) have prompted rethinking of the standard approach of convolutional networks (CNNs)~\cite{dosovitskiy2020image}. Orbital Mixer is directly based on the recently-proposed MLP Mixer vision architecture~\cite{mlpmixer}, a competitive but dramatically simpler alternative to both ViTs and CNNs. The MLP Mixer architecture splits an input image into patches and alternates between patch-wise and channel-wise mixing operations via simple multi-layer perception (MLP) layers. Though developed for vision, the MLP Mixer can be trivially adapted to other domains, and by replacing image patches with atomic orbital and overlap matrix information, we obtain a model that learns atomic interactions at any range without the need for complex or hand-engineered mixing operations. We further find that the favorable complexity of MLP layers, particularly as compared to graph transformers, to be of benefit in the molecular inference domain, where scalability to larger systems is particularly desirable.

\subsection{DFT fundamentals}

Electronic structure calculations typically represent electrons in a basis-set of atomic orbitals (AO) $\{\ket{\phi_i}\}_{i=1}^{i=N_{orbs}}$ meant to describe the available electron orbitals of the system. (Here we use the braket notation $\ket{x}$ to represent a quantum state in the complex Hilbert space.) In the Hartree-Fock (HF) model, the electron energies are given by the Hartree-Fock equations, represented in matrix form as
\begin{equation} \label{eqn:scf}
    \mathbf{F}\mathbf{C}=\mathbf{S} \mathbf{C} \mathbf{E}
\end{equation}
which determine the molecular orbital (MO) wavefunctions $\ket{\psi_m}=\sum_{i=1}^{i=N_{orbs}}C_{im}\ket{\phi_i}$ and their associated MO energies $\epsilon_m=E_{mm}$, where $\mathbf{E}$ is a diagonal matrix~\cite{lehtola2020overview}. Electrons populate the lowest energy MOs in accordance with the Pauli exclusion principle, which then define the electronic and chemical properties of the system. The Hamiltonian $\mathbf{F}$ in Eqn.~\ref{eqn:scf} approximates the single-electron energy operator $\hat{F}$ within the set of basis functions $\{\ket{\phi_i}\}$ with matrix elements defined as $F_{ij}=\bra{\phi_i}\hat{F}\ket{\phi_j}$, while the overlap matrix $\mathbf{S}$ describes the relationships between the different basis functions via the inner product $S_{ij}=\bra{\phi_i}\ket{\phi_j}$. The formulation for density-functional theory (DFT) is almost identical, with the replacement of the Kohn-Sham Hamiltonian for the Fock matrix and Kohn-Sham orbitals for molecular wavefunctions; we will use the $\mathbf{F}$ matrix or term "Hamiltonian" for either.  

The generalized eigenvalue problem in Eqn.~\ref{eqn:scf} can be solved to determine the electron density $D_{ij}=\sum_{k} C_{ik} C_{jk}$, where the summation is carried over the $k$ indexing the lowest energy MOs which are occupied. The remaining unoccupied orbitals that do not enter the density matrix calculations are called virtual orbitals and are only defined up to an arbitrary unitary transformation~\cite{schutt2019unifying}. However, the matrix elements $F_{ij}$ themselves actually depend on the electron density $\mathbf{D}$ requiring that Eqn.~\ref{eqn:scf} be solved in a self-consistent manner. Typically, DFT and HF begin with an initial guess for the electron density $\mathbf{D}$ and/or orbital occupations $\mathbf{C}$, followed by an iterative procedure that alternatingly 1) uses the density estimate $\mathbf{D}$ to calculate the Hamiltonian $\mathbf{F}$ and 2) solves Eqn.~\ref{eqn:scf} to yield an improved estimate for the electron density, until some convergence criteria is met. The major source of computational expense in DFT comes from the number of these self-consistent iterations that must be performed to obtain converged electron density estimates that may in turn be used in downstream quantum chemical calculations. 

While in principle the eigenvectors $\mathbf{C}$ and eigenvalues $\mathbf{E}$ contain the same information as the Hamiltonian $\mathbf{F}$ and the overlap matrix $\mathbf{S}$, using ML to directly predict $\mathbf{C}$ and $\mathbf{E}$ is complicated by possible state degeneracies and the coefficients being defined only up to an arbitrary phase~\cite{gastegger2020deep, schutt2019unifying}. In contrast, the Hamiltonian $\mathbf{F}$ is better behaved as a smooth function of the atomic coordinates, and combined with $\mathbf{S}$ and Eqn.~\ref{eqn:scf}, can be used to determine $\mathbf{C}$ and $\mathbf{E}$. These properties make the Hamiltonian $\mathbf{F}$ a more suitable target for ML-enabled prediction. The goal of this work is therefore to learn to reliably predict the Hamiltonian $\mathbf{F}$ for a given molecular configuration and thereby alleviating some computational expense required in self-consistently solving Eqn.~\ref{eqn:scf}.

\section{Methods} \label{sec:methods}
\subsection{Basis set-specific molecular encodings}

\begin{figure*}[htb!]
\centering
\includegraphics[width=\linewidth]{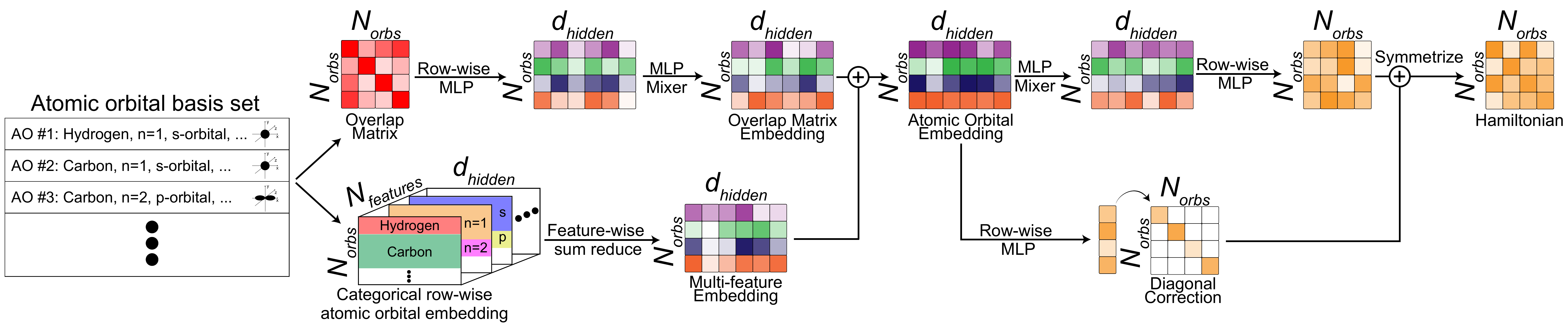}
\caption{Orbital Mixer neural network architecture for predicting molecular electronic structure. Orbital Mixer leverages properties of atomic orbitals that comprise the DFT basis set to predict the Hamiltonian $\mathbf{F}$ in Eqn.~\ref{eqn:scf}, from which the molecular orbital coefficients defining the electron density are derived.}
\label{fig:arch}
\end{figure*}

Many existing deep learning architectures for molecular inference use exclusively basis-independent information, even when predicting basis-dependent information (e.g., the Hamiltonian $\mathbf{F}$). In contrast, we explicitly choose a basis that the network will model, and represent each molecular configuration through basis-dependent quantities. This featurization approach is similar OrbNet~\cite{qiao2020orbnet,qiao2020multi}, which uses symmetry-adapted atomic orbital features within a GNN architecture for molecular property prediction. In particular, we use the overlap matrix $\mathbf{S}$ to capture geometric information together with categorical features corresponding to each atomic orbital. Elements of the overlap matrix $S_{ij}$ measure the spatial overlap of localized atomic orbital basis functions, and provide a detailed description of the molecular geometry specific to the choice of basis set. Unlike the Hamiltonian $\mathbf{F}$, overlap matrix elements are independent of electron density, and can be quickly computed given only a choice of basis set and the atomic positions.

Though supplying the model with atomic coordinates directly as inputs provides benefits -- most notably, native access to analytic derivatives with respect to the atomic positions, which may be later used for e.g., electronic property optimization~\cite{schutt2019unifying} -- exposing basis set information directly as inputs to Orbital Mixer provides a stronger inductive bias than basis-independent features like atomic coordinates, and gives a more direct link when predicting the Hamiltonian $\mathbf{F}$ that is expressed within the same basis set and carries the same symmetries as the overlap matrix $\mathbf{S}$. These representations are then combined with categorical encodings of atomic orbital features that comprise the basis set. Together, the overlap matrix $\mathbf{S}$ and categorical atomic orbital features provide detailed geometric and basis set-specific descriptors for us to effectively predict the molecular electronic structure.    

%\noter{Thoughts on including this paragraph below? Not sure if necessary to even discuss this directly and would be okay with removing -- thought it seemed most relevant to the points we mention in this section so I put it here.}

%An appeal of the SchNOrb and PhiSNet framework is that they accept atomic coordinates directly as input which enables native accessibility to analytic derivatives with respect to the atomic positions. As demonstrated by SchNOrb, this property lends itself to applications in explainability and electronic property optimization~\cite{schutt2019unifying}. In our model we use the overlap matrix $\mathbf{S}$ as a representation of the molecular configuration, which given a predefined basis, is function of only the atomic coordinates and in principle also grants access to analytic derivates with respect to the atomic positions. Access to such derivatives have been studied in detail for the symmetry adapted atomic orbital representations used as input for the OrbNet architecture~\cite{qiao2020multi}. Explorations of these analytic derivates for our model are out of the scope of this paper and left to future work.

\subsection{The architecture}
Orbital Mixer presents a novel MLP mixer-based architecture for modelling molecular electronic structure in a basis of localized atomic orbitals. A schematic illustration of the Orbital Mixer deep learning architecture and data flow is presented in Fig.~\ref{fig:arch}. We begin with enumerating a list of atomic orbitals comprising the set of basis functions used to model the electronic Hartree-Fock or Kohn-Sham wavefunction of the system. The size and complexity of the basis set determines the achievable accuracy of the electronic structure calculation at the cost of computational expense, and depends on the particular system and properties of interest. Each atomic orbital is characterized with five categorical features that uniquely specify each orbital within the basis set: the index and element of the atom at which the orbital is centered, the principal quantum number \textit{n}, the azimuthal quantum number \textit{l}, and the magnetic quantum number \textit{m$_l$}. Each feature for each orbital is transformed into a $d_{\text{hidden}}$-dimensional vector using a separate learned embedding layer and assembled into a tensor of shape $N_{\text{orbs}} \times d_{\text{hidden}} \times N_{\text{features}}$. Finally performing a summation over the $N_{\text{features}}$ dimensions yields the complete multi-feature embedding for the atomic orbital basis of shape $N_{\text{orbs}} \times d_{\text{hidden}}$.

While the multi-feature embedding describes properties of each atomic orbital independent of the molecular geometry, the overlap matrix $\mathbf{S} \in \mathbb{R}^{N_{\text{orbs}} \times N_{\text{orbs}}}$ provides a representation of the molecular geometry by measuring the integrable overlap of the atomic orbital functions basis which depends on the spatial arrangement of the atoms where the orbitals are localized. Each row $i$ of $S$ corresponds to the overlap of an atomic orbital $\ket{\phi_i}$ with the other $N_{\text{orbs}}$ orbitals; these rows are processed with an initial MLP to yield $d_{\text{hidden}}$-dimensional atomic orbital representations, and their interactions are modeled with subsequent MLP Mixer layers. As the overlap matrix $\mathbf{S}$ is covariant with respect to rigid rotations of the atomic coordinates, this processing step is important to capture the interdependence between the atomic orbitals and learn a globally aware overlap matrix embedding of shape $N_{\text{orbs}} \times d_{\text{hidden}}$. The multi-feature embedding is then added to this overlap matrix embedding yielding a complete atomic orbital embedding that captures both categorical properties specific to each atomic orbital along with global spatial information of the molecular geometry.            

The atomic orbital representations are further refined with another series of MLP mixers, followed by a row-wise MLP to reshape the $N_{\text{orbs}} \times d_{\text{hidden}}$ representations into the target $N_{\text{orbs}} \times N_{\text{orbs}}$ dimensionality of the Hamiltonian $\mathbf{F}$. As the off-diagonal elements of the Hamiltonian $F_{i,j}$ are indicative of coupling between the atomic orbitals, the MLP mixer layer effectively captures these interactions between atomic orbitals useful for predicting the off-diagonal matrix elements. On the other hand, the diagonal elements of the Hamiltonian $F_{i,i}$ correspond to energies of each of the atomic orbital basis functions, and in practice are often significantly larger than their off-diagonal counterparts in magnitude. To account for these physical and numerical differences, we apply a separate MLP row-wise to the atomic orbital embeddings generating a single scalar value for each of the $N_{\text{orbs}}$ atomic orbitals. This $N_{\text{orbs}}$-dimensional vector is assembled into a diagonal matrix and added to the $N_{\text{orbs}} \times N_{\text{orbs}}$ dimensional output of the interaction branch, constituting a diagonal correction to the interaction branch and providing an inductive bias delineating the on- and off-diagonal elements of the Hamiltonian $\mathbf{F}$. Finally, as the Hamiltonian $\mathbf{F}$ is always symmetric, the diagonally corrected representations $\Tilde{\mathbf{F}}$ are symmetrized to obtain the complete predicted Hamiltonian $\mathbf{F} = \frac{1}{2} (\mathbf{\Tilde{F}} + \mathbf{\Tilde{F}^T)}$. From the predicted Hamiltonian $\mathbf{F}$ we can obtain the molecular orbital coefficients, energies and electron density by solving Eqn.~\ref{eqn:scf}, giving us access to the molecular electronic structure and a plethora of chemical and physical properties for the system.

\subsection{Training procedures}

Orbital Mixer is trained end-to-end with mini-batch stochastic gradient descent and the ADAM optimizer~\cite{kingma2014adam} using a simple mean squared error (MSE) between the true $\mathbf{F}^{(\text{true})}$ and predicted $\mathbf{F}^{(\text{pred})}$ Hamiltonians,
\begin{equation}
    \mathcal{L}(\mathbf{F}^{(\text{true})}, \mathbf{F}^{(\text{pred})}) = \frac{1}{N_{\text{orbs}}^2} \|\mathbf{F}^{(\text{true})} - \mathbf{F}^{(\text{pred})}\|_F^2
\end{equation}
Importantly, both the input overlap matrix $\mathbf{S}$ and the Hamiltonian $\mathbf{F}$ are covariant with respect to rigid rotations of the atomic coordinates $\mathbf{r} \in \mathbb{R}^{N_{\text{atoms}} \times 3}$. We perform data augmentation to ensure Orbital Mixer leans this covariance where during training a random rotation matrix $\mathbf{R}$ performs a rigid rotation of the atomic coordinates $\mathbf{r}^{\prime} = \mathbf{r} \mathbf{R}^T$ for each training sample. The corresponding covariant change to the overlap $\mathbf{S}$ and Hamiltonian $\mathbf{F}$ matrices due to this rigid rotation $\mathbf{R}$ is then accounted for using Wigner $\mathcal{D}_{\mathbf{R}} \in \mathbb{R}^{N_{\text{orbs}} \times N_{\text{orbs}}}$ rotation matrices~\cite{wigner1931gruppentheorie} via a unitary transformation,
\begin{align*}
    \mathbf{S}^{\prime} &= \mathcal{D}_{\mathbf{R}}^{\text{T}} \mathbf{S} \mathcal{D}_{\mathbf{R}} \\
    \mathbf{F}^{\prime} &= \mathcal{D}_{\mathbf{R}}^{\text{T}} \mathbf{F} \mathcal{D}_{\mathbf{R}}.
\end{align*}

We also incorporate during training a separately maintained exponential moving average of the model parameters that are then used at inference time, as we find this leads to improved generalizability. Orbital Mixer in total contains $\sim$38M parameters, which is $\sim$2.5$\times$ fewer than the $\sim$93M parameters in the SchNOrb architecture, but $\sim$2$\times$ more than the $\sim$17M parameters of PhiSNet. A full accounting of all neural network, optimizer and scheduler hyperparameters is provided in the Appendix along with a complete PyTorch~\cite{paszke2019pytorch} implementation using PyTorch Lightning~\cite{falcon2019pytorch} publicly available at DOI:10.18126/cu4h-d2mm~\cite{MDF}.

\section{Results}

\subsection{Predicting molecular electronic structure}
The neural network architecture proposed in this work, Orbital Mixer, is capable of accurately predicting the Hamiltonian $\mathbf{F}$ used to determine the electronic density and other derivable physical and chemical properties for a variety of molecular systems. We evaluate Orbital Mixer on conformational geometries of small molecule molecular dynamics trajectories taken from the MD17 dataset~\cite{chmiela2017machine}. Namely, we select the three molecules (Ethanol, Malondialdehyde and Uracil) from the MD17 dataset that were investigated in the original SchNOrb paper. While the MD17 dataset natively contains only energy and force labels, we curate our dataset by performing reference DFT calculations using the PySCF~\cite{sun2018pyscf} quantum chemistry code on the same subset of $\sim$30,000 MD17 molecular configurations for each molecule used in SchNOrb and PhisNet~\cite{schutt2019unifying}. We train separate neural networks for each molecule at two different training set sizes of 25K and 950 configurations alongside comparisons to results from SchNOrb and PhiSNet. Complete details for all training settings are provided in the Appendix.

%We evaluate our method against the baseline performance of SchNOrb, which was originally only trained on 25K configurations. Using the publically available SchNOrb implementation \noter{Should we provide a github link to this implimentation here?} we reproduce the published results to within $\sim$5\% for the 25K training size \noter{Do we need to show evidence that we reproduce to within 5\% or sufficent to just state it? Do we even need to make this statement?}, while also training separate SchNOrb networks at a training size of 950 for each molecule. Complete details for all training settings are provided in the Appendix \noter{Here I'm referencing Table 3 and Appendix section 3, which is the same Appendix section I reference in Section 3.4 -- is this okay?}.      

Numerical results of Orbital Mixer's performance compared to SchNOrb and PhiSNet are presented in Table~\ref{tab:1}. For each tested molecule and training size, Orbital Mixer outperforms SchNOrb and achieves mean absolute errors (MAEs) below 0.003 eV on all Hamiltonian predictions, but fails to outperform PhiSNet, which greatly benefits from the built-in covariance with respect to rigid rotations of their SE(3)-equivariant architecture. This is consistent with previous work where incorporating built-in covariance improves model prediction \cite{Park2021,batzner2021se}. Impressively, however, Orbital Mixer trained on only 950 configurations for each test molecule generates more accurate Hamiltonian $\mathbf{F}$ MAE predictions than SchNOrb trained with 25K configurations. We notice the greatest improvement compared to SchNOrb when training on the largest and most challenging molecule, uracil, which is modeled in a basis set of 132 atomic orbitals (29 occupied + 103 virtual). Compared to a SchNOrb model trained with 25K samples, we achieve $\sim$53\% improvement for Orbital Mixer trained with 950 samples and $\sim$61\% improvement when trained with 25K samples on Uracil, but still performs $\sim$158\% worse than PhiSNet in this setting. Orbital Mixer achieves impressive accuracy on occupied MO energy and coefficient prediction, outperforming SchNOrb on these metrics while predicting MO energies to within 0.0075 eV MAE and MO coefficients to greater than 99\% cosine similarity for all test molecules and training set sizes. The better Hamiltonian $\mathbf{F}$ prediction accuracy of PhiSNet ultimately leads to improved predictions errors for MO energies as well, and on uracil performs $\sim$88\% better than Orbital Mixer.% However, the data efficiency of PhiSNet, i.e. error when training on smaller data sets, is unknown.

\begin{table}[htb!]
\centering
\resizebox{0.5\textwidth}{!}{%
\begin{tabular}{@{}ccccc@{}}
\toprule
Molecule                         & \begin{tabular}[c]{@{}c@{}}Model \\ (train size)\end{tabular}  & \begin{tabular}[c]{@{}c@{}}Hamiltonian \\ MAE {[}eV{]}\end{tabular} & \begin{tabular}[c]{@{}c@{}}MO energy \\ MAE {[}eV{]}\end{tabular} & \begin{tabular}[c]{@{}c@{}}MO coefficient \\ cosine similarity\end{tabular} \\ \midrule
\multirow{4}{*}{Ethanol}         & Orbital Mixer (25K) & 0.0020                                                     & 0.0037                                                   & \textbf{0.9999}                                                             \\ \cmidrule(l){2-5} 
                                 & \textbf{PhiSNet (25K)}       & \textbf{0.00033}                                                              & \textbf{0.0017}                                                            & \textbf{1.00}\footnote[1]{}                                                                      \\ \cmidrule(l){2-5} 
                                 & SchNOrb (25K)       & 0.0052                                                              & 0.0084                                                            & 0.9978                                                                      \\ \cmidrule(l){2-5} 
                                 & \textbf{Orbital Mixer (950)} & \textbf{0.0026}                                                     & \textbf{0.0054}                                                   & \textbf{0.9998}                                                             \\ \cmidrule(l){2-5} 
                                 & SchNOrb (950)       & 0.0074                                                              & 0.0130                                                            & 0.9941                                                                      \\ \midrule
\multirow{4}{*}{Malondialdehyde} & Orbital Mixer (25K) & 0.0021                                                     & 0.0046                                                  & \textbf{0.9984}                                                             \\ \cmidrule(l){2-5} 
                                 & \textbf{PhiSNet (25K)}       & \textbf{0.00034}                                                              & \textbf{0.0020}                                                            & \textbf{1.00}\footnote[1]{}                                                                      \\ \cmidrule(l){2-5} 
                                 & SchNOrb (25K)       & 0.0052                                                              & 0.0117                                                            & 0.9866                                                                      \\ \cmidrule(l){2-5} 
                                 & \textbf{Orbital Mixer (950)} & \textbf{0.0029}                                                     & \textbf{0.0064}                                                   & \textbf{0.9973}                                                             \\ \cmidrule(l){2-5} 
                                 & SchNOrb (950)       & 0.0075                                                              & 0.0221                                                            & 0.9661                                                                      \\ \midrule
\multirow{4}{*}{Uracil}          & Orbital Mixer (25K) & 0.0025                                                     & 0.0059                                               & \textbf{0.9965}                                                             \\ \cmidrule(l){2-5} 
                                 & \textbf{PhiSNet (25K)}       & \textbf{0.00029}                                                              & \textbf{0.0023}                                                            & \textbf{1.00}\footnote[1]{}                                                                      \\ \cmidrule(l){2-5} 
                                 & SchNOrb (25K)       & 0.0064                                                              & 0.0355                                                            & 0.9269                                                                      \\ \cmidrule(l){2-5} 
                                 & \textbf{Orbital Mixer (950)} & \textbf{0.0030}                                                     & \textbf{0.0074}                                                   & \textbf{0.9941}                                                             \\ \cmidrule(l){2-5} 
                                 & SchNOrb (950)       & 0.0086                                                              & 0.1550                                                            & 0.8003                                                                      \\ \bottomrule
\end{tabular}%
}
\caption{Comparison of Hamiltonian $\mathbf{F}$, occupied molecular orbital (MO) energies and MO coefficient prediction accuracies between Orbital Mixer, SchNOrb~\cite{schutt2019unifying} and PhiSNet~\cite{unke2021se} for the three test molecules of Ethanol, Malondialdehyde and Uracil. We generate comparisons using two different training set sizes of 950 and 25K configurations for both Orbital Mixer and SchNOrb and compare to the reported results at 25K training samples for PhiSNet\footnote[2]{}. Best results for each molecule and training set size are shown in bold.}
\label{tab:1}
\end{table}
\footnotetext[1]{Reported PhiSNet cosine similarities are rounded to fewer significant digits than we report in this work.}
\footnotetext[2]{Results for PhiSNet are taken from Ref.~\cite{unke2021se} and SchNOrb models are trained in this work using the publically available implementation at: https://github.com/atomistic-machine-learning/SchNOrb}
% Model (train size)

\begin{figure*}[htb!]
\centering
\includegraphics[width=\linewidth]{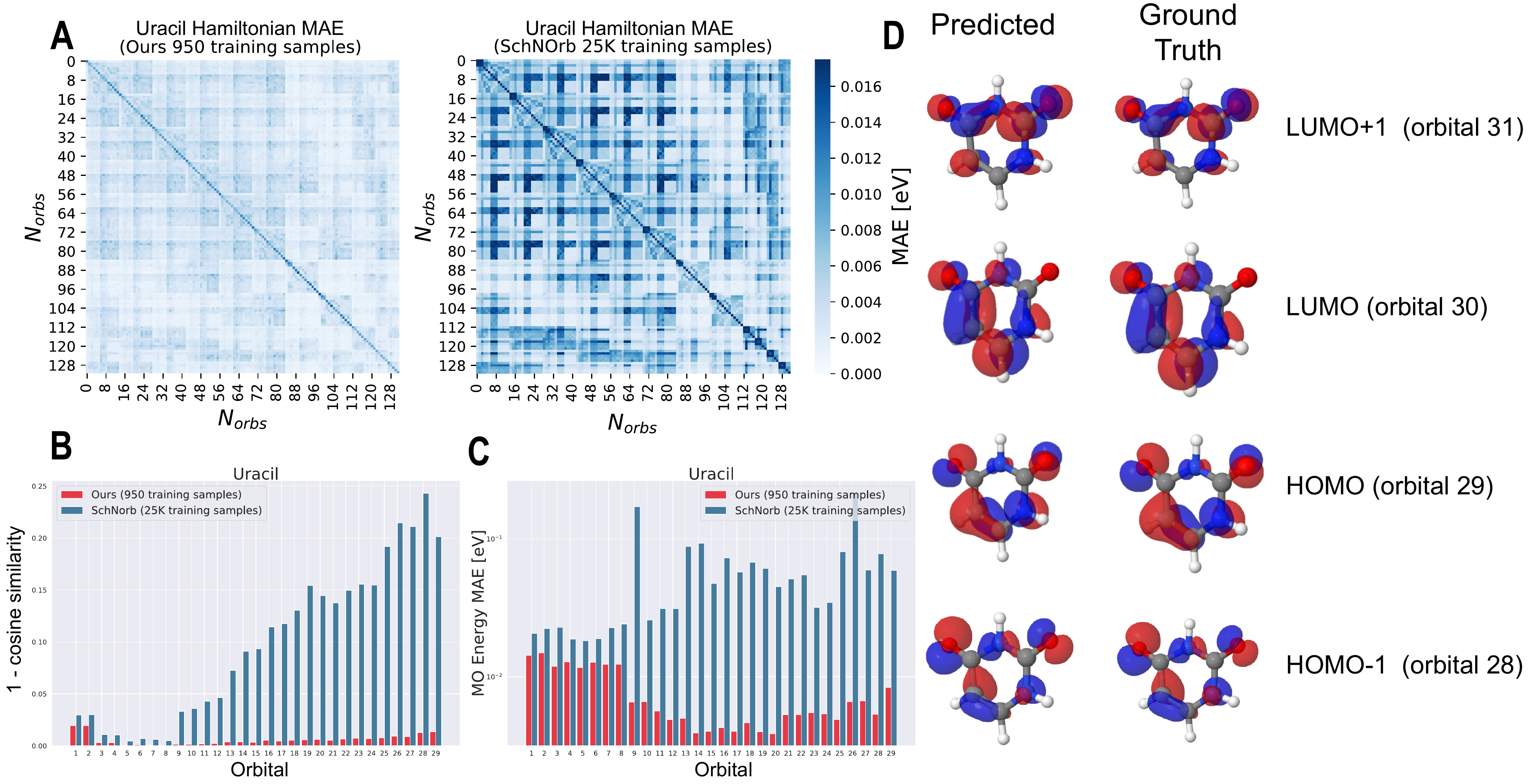}
\caption{Prediction errors for Uracil of \textbf{(a)} the Hamiltonian $\mathbf{F}$, \textbf{(b)} the molecular orbital (MO) coefficients and \textbf{(c)} the MO energy between Orbital Mixer predictions when trained with 950 configurations and SchNOrb trained with 25K configurations. \textbf{(d)} Visualization of Uracil molecular orbital shapes derived from Orbital Mixer predicted and ground truth MO coefficients.}
\label{fig:viz}
\end{figure*}

In Fig.~\ref{fig:viz} we present a detailed comparison of Hamiltonian $\mathbf{F}$ and MO coefficients and energies predictions generated by Orbital Mixer trained with 950 configurations and SchNOrb trained with 25K configurations. We visualize in Fig.~\ref{fig:viz}A the matrix element-wise test set MAE of the predicted Uracil Hamiltonian $\mathbf{F}$. Beyond the concentration of both models' prediction errors along the diagonal (as expected, due to the diagonal entries' significantly larger magnitude), SchNOrb model predictions produce significantly larger errors in select off-diagonal blocks, while errors from Orbital Mixer predictions are comparatively lower throughout the off-diagonal elements. In Fig.~\ref{fig:viz}B we show cosine similarity between SchNOrb and Orbital Mixer predicted MO coefficients delineated for the 29 occupied orbitals of Uracil. Orbital Mixer never performs worse than $\sim$0.983 cosine similarity for any particular occupied orbital, while SchNOrb only performs better than our worst performer for 6/29 occupied orbitals. We also show in Fig.~\ref{fig:viz}C the MAE between the ground truth and predicted occupied MO energies for both Orbital Mixer and SchNOrb. Here we notice our largest error orbital (orbital 2 at $\sim$0.012 eV) performs better than all the predicted SchNOrb orbitals and $\sim$52\% better than the best SchNOrb orbital. We emphasize Orbital Mixer achieves these predictions accuracies using $\sim$10$\times$ fewer training samples than SchNOrb, but nevertheless on average performs worse than PhiSNet (Table~\ref{tab:1}). 

Fig.~\ref{fig:viz}D depicts the shapes of the frontier molecular orbitals for a fixed Uracil configuration derived from both Orbital Mixer predicted and ground truth MO coefficients. The orbital shapes produced by Orbital Mixer predictions are visually identical to the ground truth for both the Highest Occupied Molecular Orbital (HOMO), the Lowest Unoccupied Molecular Orbital (LUMO) and the two nearby occupied (HOMO-1) and unoccupied (LUMO+1) molecular orbitals. Similar figures displaying errors for the Ethanol and Malondialdehyde datasets along with visualizations of their orbital shapes is provided in the Appendix.

\subsection{Improved SCF convergence}

\begin{figure}[htb!]
\centering
\includegraphics[width=\linewidth]{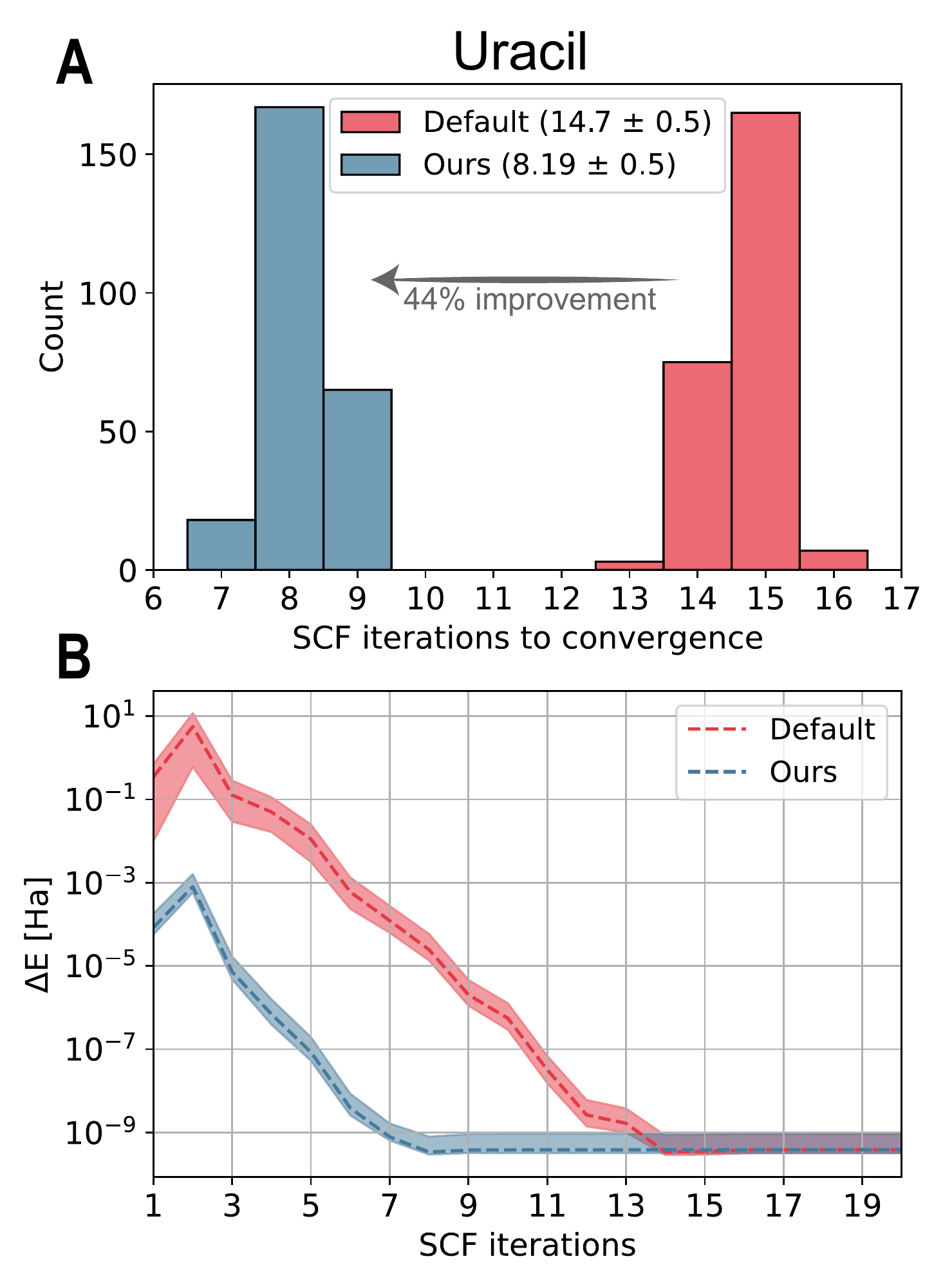}
\caption{\textbf{(a)} Number of SCF iterations required to reach convergence using initial guesses for the Hamiltonian $\mathbf{F}$ generated by Orbital Mixer and the default PySCF initialization for 250 test set Uracil configurations. \textbf{(b)} Difference in the total energy at each SCF iterations compared to the terminal converged energy estimate. Errors about the mean dotted line bound the 25\% and 75\% quartiles.}
\label{fig:converge}
\end{figure}

The ability to accurately predict the Hamiltonian $\mathbf{F}$ for a molecular system enables integration of Orbital Mixer with electronic structure calculations. Namely, the Hamiltonian predictions generated by Orbital Mixer can be used as initial guess in Eqn.~\ref{eqn:scf}, which is then solved using the self-consistent field (SCF) method to arrive at a converged estimate of the Hamiltonian $\mathbf{F}$, and therefore the electron density. The speed of these DFT calculations is in large part determined by the number SCF iterations required to reach convergence. High quality initial guesses for the Hamiltonian $\mathbf{F}$ can greatly expedite the SCF procedure and enable higher-throughput DFT calculations. We perform experiments testing speed-up with respect to number of SCF iterations by initializing DFT calculations using Orbital Mixer predicted Hamiltonian $\mathbf{F}$. Reported in Fig.~\ref{fig:converge}A is the distribution of the SCF iterations to convergence for DFT calculations performed on 250 test set Uracil configurations using both the default PySCF and Orbital Mixer predicted Hamiltonian $\mathbf{F}$ for initialization. The Orbital Mixer initialization achieves an impressive $\sim$44\% improvement in the number of SCF iterations required to reach convergence compared to the default PySCF initialization strategy. Fig.~\ref{fig:converge}B visualizes this improvement by tracking the difference in the total energy estimated after each SCF iteration as compared to the terminal converged energy estimate. Although both initialization schemes eventually reach the same energy difference criterion of 10$^{-9}$ Ha, all configurations using the Orbital Mixer initialization reliably converge after only at most 9 SCF iterations. In similar SCF initialization experiments, SchNorb reports a speedup of $\sim$15\% and PhiSNet a speedup of $\sim$47\% SCF iterations for Uracil. Similar comparisons for SCF convergence improvement of Orbital Mixer for the Ethanol and Malondialdehyde datasets are provided in the Appendix. We expect similar performance for other self-consistent field applications, from higher-accuracy quantum chemistry methods to inhomogeneous copolymers and nanoparticles \cite{Arora2016}.      

\subsection{Derived physical chemical properties}

%Recent research has focused on developing ML models trained to separately predict select molecular properties, such as spectroscopic properties, optoelectronic properties and activation energies. Directly predicting the Hamiltonian $\mathbf{F}$ for a molecular configuration defines the electronic structure via the molecular orbital coefficients, which in turn provides access to a plethora of derived physical and chemical properties without the need to train separate models for each property of interest.

We interrogate the ability of Orbital Mixer to directly calculate physical and chemical properties by using Orbital Mixer predicted Hamiltonian $\mathbf{F}$ and molecular orbital coefficients to calculate HOMO-LUMO gaps and electronic dipole moments, both of which are physically meaningful and often measurable quantities. Reported in Table.~\ref{tab:2} is a comparison of predicted HOMO-LUMO gap energies and dipole moments between Orbital Mixer and SchNOrb; we note that dipole moments or HOMO-LUMO gap energies were not reported in the original PhiSNet paper. The results for Orbital Mixer show overall excellent accuracies with respect to the reference DFT calculations for both HOMO-LUMO gap energies and dipole moments, at both training sizes MAEs below than 0.017 eV and 0.034 D, respectively. We again outperform SchNOrb using $\sim$10$\times$ fewer training data on all three benchmark molecules. Orbital Mixer performs particularly well on the most challenging molecule Uracil: using only 950 training samples, we generate HOMO-LUMO gap energy and dipole moment predictions with errors orders of magnitude smaller than those from a SchNOrb model trained with 25K configurations. These results highlight the ability of Orbital Mixer to capture chemically and physically meaningful molecular properties using a single neural network architecture, circumventing the need to develop separate specialized models for each property of interest.

\begin{table}[htb!]
\centering
\resizebox{0.5\textwidth}{!}{%
\begin{tabular}{@{}cccc@{}}
\toprule
Molecule                         & \begin{tabular}[c]{@{}c@{}}Model \\ (train size)\end{tabular} & \begin{tabular}[c]{@{}c@{}}HOMO-LUMO\\ gap {[}eV{]}\end{tabular} & \begin{tabular}[c]{@{}c@{}}Dipole \\ moment {[}D{]}\end{tabular} \\ \midrule
\multirow{4}{*}{Ethanol}         & \textbf{Orbital Mixer (25K)}                                           & \textbf{0.0115}                                                  & \textbf{0.0071}                                                  \\ \cmidrule(l){2-4} 
                                 & SchNOrb (25K)                                                 & 0.0743                                                           & 0.0262                                                           \\ \cmidrule(l){2-4} 
                                 & \textbf{Orbital Mixer (950)}                                           & \textbf{0.0163}                                                  & 0.0103                                                           \\ \cmidrule(l){2-4} 
                                 & SchNOrb (950)                                                 & 0.1190                                                           & --                                                               \\ \midrule
\multirow{4}{*}{Malondialdehyde} & \textbf{Orbital Mixer (25K)}                                           & \textbf{0.0061}                                                  & \textbf{0.0132}                                                  \\ \cmidrule(l){2-4} 
                                 & SchNOrb (25K)                                                 & 0.0384                                                           & 0.0536                                                           \\ \cmidrule(l){2-4} 
                                 & \textbf{Orbital Mixer (950)}                                           & \textbf{0.0083}                                                  & 0.0187                                                           \\ \cmidrule(l){2-4} 
                                 & SchNOrb (950)                                                 & 0.1239                                                           & --                                                               \\ \midrule
\multirow{4}{*}{Uracil}          & \textbf{Orbital Mixer (25K)}                                           & \textbf{0.0074}                                                  & \textbf{0.0227}                                                  \\ \cmidrule(l){2-4} 
                                 & SchNOrb (25K)                                                 & 0.4503                                                           & 1.2762                                                           \\ \cmidrule(l){2-4} 
                                 & \textbf{Orbital Mixer (950)}                                           & \textbf{0.0119}                                                  & 0.0336                                                           \\ \cmidrule(l){2-4} 
                                 & SchNOrb (950)                                                 & 1.2780                                                           & --                                                               \\ \bottomrule
\end{tabular}%
}
\caption{Comparison of derived physical chemical properties between Orbital Mixer and SchNOrb for each tested molecule at different training sizes. The HOMO-LUMO gaps and dipole moments are calculated with PySCF using Orbital Mixer predicted Hamiltonian $\mathbf{F}$. The HOMO-LUMO gap was not reported in the original SchNOrb paper and are therefore calculated using our retrained SchNOrb models. We report dipole moments for SchNOrb only at the 25K training set sizes that were reported in the original paper.}
\label{tab:2}
\end{table}

\section{Conclusions}

We present in this work a deep learning workflow for predicting molecular electronic structure directly in a basis of localized atomic orbitals. Compared to SchNOrb and PhiSNet, which generate predictions as a function of only atomic coordinates and molecular composition alone, Orbital Mixer leverages strong inductive biases by operating jointly on basis set-specific atomic orbital representations and the overlap matrix $\mathbf{S}$ to predict the Hamiltonian $\mathbf{F}$ for a molecular configuration. However, unlike PhiSNet which achieves explicit covariance using built-in SE(3)-equivariant operations, Orbital Mixer implicitly learns covariance similar to SchNOrb by training with data augmentation. Orbital Mixer benefits from a simple and intuitive architecture modelling interactions between atomic orbital representations using MLP mixers. Compared to SchNOrb, we achieve upwards of 50\% improvement in Hamiltonian $\mathbf{F}$ mean absolute errors and upwards of 95\% improvement in predicting derived physical chemical properties while using $\sim$10$\times$ fewer training samples. Nevertheless the built-in covariance of PhiSNet with respect to rigid molecular rotations and translations proves to be invaluable for prediction accuracy and therefore enables PhiSNet to achieve $\sim$158\% better reported Hamiltonian $\mathbf{F}$ MAE. We also demonstrate how integrating Orbital Mixer into DFT workflows by purposing Orbital Mixer predicted Hamiltonian $\mathbf{F}$ as an initial guess yields $\sim$44\% improvement in the number of SCF iterations required to reach convergence for Uracil. Orbital Mixer scales well from the smallest molecule in our dataset, Ethanol, containing only 72 atomic orbitals, to the most challenging molecule, Uracil, with 132 atomic orbitals. This work represents an alternate approach for molecular electronic structure prediction leveraging a novel basis set dependent featurization within a simple MLP Mixer-enabled deep learning architecture.

%This work represents an important step towards incorporating ML into the quantum chemistry workflows. \noter{Need more here on future outlook for this work}  

\section*{Data Availability}
A complete PyTorch implementation of Orbital Mixer alongside evaluation Jupyter notebooks and all data are available via the Materials Data Facility (MDF)~\cite{blaiszik2016materials,blaiszik2019data} at DOI:10.18126/cu4h-d2mm~\cite{MDF}.  

\section*{Acknowledgements}
KS is supported by the National Science Foundation Graduate Research Fellowship under Grant No. DGE-1746045.


\bibliography{main}
\bibliographystyle{icml2021}

\appendix
\section{DFT calculation details}
We evaluate Orbital Mixer on reference DFT calculations performed on Ethanol, Malondialdehyde and Uracil molecule configurations extracted from the MD17 dataset~\cite{chmiela2017machine}. Reference DFT calculations were performed on a subset of configurations that were used for training and evaluation in the SchNOrb paper~\cite{schutt2019unifying}. We replicate the DFT calculations outlined in SchNOrb using the PySCF quantum chemistry code~\cite{sun2018pyscf} to generate our datasets. The def2-SVP basis set~\cite{weigend2005balanced} was used with the PBE exchange correlation functional~\cite{perdew1996generalized}. All default PySCF procedures were used for performing SCF iterations based on the direct inversion in the iterative subspace (DIIS) method~\cite{pulay1980convergence,pulay1982improved} with default initial guesses generated using the `MinAO' method~\cite{sun2018pyscf} that considers a superposition of atomic densities projected onto the first contracted functions in the cc-pVTZ or cc-pVTZ-PP basis set. Each calculation uses a convergence criterion of 10$^{-13}$ Ha total energy difference between consecutive iterations or a maximum of 50 SCF iterations.

\section{Neural network architecture and training settings}

Throughout the Orbital Mixer architecture we use a hidden dimension of $d_{\text{hidden}}=1024$ and GELU activation functions~\cite{hendrycks2016gaussian}. The initial row-wise MLP used to process the rows of the overlap matrix $\mathbf{S}$ includes two dense layers with an expansion factor of 2 for the intermediate hidden representation, such that the complete action of the MLP involves the following sequence of transformations onto the shape of the atomic orbital representations: $N_{\text{orbs}} \rightarrow 2d_{\text{hidden}} \xrightarrow[]{\text{GELU}} d_{\text{hidden}}$. The MLPs within the MLP Mixer layers for the overlap matrix and interaction branch similarly use an expansion factor of 2 with no dropout. The initial MLP Mixer applied to the overlap matrix $\mathbf{S}$ consists of $n_{\text{layers}}=2$ Mixer layers, while the second MLP Mixer applied to the atomic orbital embeddings used for predicting the Hamiltonian $\mathbf{F}$ within the interaction branch uses $n_{\text{layers}}=6$ Mixer layer. The row-wise MLP used to reshape the $N_{\text{orbs}} \times d_{\text{hidden}}$ representation processed by the MLP Mixer in the interaction branch into the target $N_{\text{orbs}} \times N_{\text{orbs}}$ dimensionality consists of a GELU non-linearity followed by a single dense layer. The row-wise MLP in the diagonal correction branch similarly uses a GELU non-linearity followed by a single dense layer. A complete PyTorch implementation of Orbital Mixer with accompanying training scripts is publicly available at DOI:10.18126/cu4h-d2mm~\cite{MDF}.

Orbital Mixer is trained using the ADAM optimizer with the default PyTorch parameters and a mini-batch size of 32 samples. Evaluation and testing are performed using the model obtained with an exponential moving average over all parameters during training time employing a decay rate of 0.999 per step. Gradient clipping is applied to clip gradient norms to a maximum value of 0.001, as we find this helps to stabilize training. We use an initial learning rate of $3 \times 10 ^{-4}$ which we decay by a factor of $\gamma=0.8$ every $n_{\text{decay}}$ training steps. All Orbital Mixer models are trained for 120 hours on Nvidia Tesla V100 32GB GPUs, after which time we observe the training and validation losses to plateau. When training reference SchNOrb models we follow training procedures outlined in the SchNOrb paper for the 25K dataset originally handled in the paper. We use all the same training settings when training SchNOrb models on the 950 sample training set size, except we modify the patience to decay the learning rate from the original 15 epochs when training with 25K samples to 150 epochs when training with 950 samples. Training of all SchNOrb models was still stopped in each case once the learning rate dropped below the $5 \times 10 ^{-6}$ threshold. Provided in Table~\ref{tab:model_param} is a breakdown of all these training settings and parameters for both Orbital Mixer and retrained SchNOrb models.

\begin{table*}[]
\centering
\resizebox{\textwidth}{!}{%
\begin{tabular}{@{}cccclll@{}}
\toprule
Molecule                         & \begin{tabular}[c]{@{}c@{}}Model \\ (train size)\end{tabular} & \begin{tabular}[c]{@{}c@{}}Val\\ size\end{tabular} & \begin{tabular}[c]{@{}c@{}}Test \\ size\end{tabular} & \multicolumn{1}{c}{\begin{tabular}[c]{@{}c@{}}Batch\\ size\end{tabular}} & \multicolumn{1}{c}{\begin{tabular}[c]{@{}c@{}}Initial\\ learning rate\end{tabular}} & \multicolumn{1}{c}{Scheduler}                                                                                \\ \midrule
\multirow{4}{*}{Ethanol}         & Orbital Mixer (25K)                                                    & 50                                                 & 4500                                                 & 32                                                                       & $3 \times 10^{-4}$                                                                  & \begin{tabular}[c]{@{}l@{}}Decay LR by $\gamma=0.8$ every\\ $n_{\text{decay}}=1M$ steps\end{tabular}         \\ \cmidrule(l){2-7} 
                                 & SchNOrb (25K)                                                 & 500                                                & 4500                                                 & 32                                                                       & $1 \times 10^{-4}$                                                                  & \begin{tabular}[c]{@{}l@{}}Decay LR by $\gamma=0.8$ after\\ 15 epochs w/o val loss improvement\end{tabular}  \\ \cmidrule(l){2-7} 
                                 & Orbital Mixer (950)                                                    & 50                                                 & 4500                                                 & 32                                                                       & $3 \times 10^{-4}$                                                                  & \begin{tabular}[c]{@{}l@{}}Decay LR by $\gamma=0.8$ every\\ $n_{\text{decay}}=500K$ steps\end{tabular}       \\ \cmidrule(l){2-7} 
                                 & SchNOrb (950)                                                 & 50                                                 & 4500                                                 & 32                                                                       & $1 \times 10^{-4}$                                                                  & \begin{tabular}[c]{@{}l@{}}Decay LR by $\gamma=0.8$ after\\ 150 epochs w/o val loss improvement\end{tabular} \\ \midrule
\multirow{4}{*}{Malondialdehyde} & Orbital Mixer (25K)                                                    & 50                                                 & 1478                                                 & 32                                                                       & $3 \times 10^{-4}$                                                                  & \begin{tabular}[c]{@{}l@{}}Decay LR by $\gamma=0.8$ every\\ $n_{\text{decay}}=1M$ steps\end{tabular}         \\ \cmidrule(l){2-7} 
                                 & SchNOrb (25K)                                                 & 500                                                & 1478                                                 & 32                                                                       & $1 \times 10^{-4}$                                                                  & \begin{tabular}[c]{@{}l@{}}Decay LR by $\gamma=0.8$ after\\ 15 epochs w/o val loss improvement\end{tabular}  \\ \cmidrule(l){2-7} 
                                 & Orbital Mixer (950)                                                    & 50                                                 & 1478                                                 & 32                                                                       & $3 \times 10^{-4}$                                                                  & \begin{tabular}[c]{@{}l@{}}Decay LR by $\gamma=0.8$ every\\ $n_{\text{decay}}=300K$ steps\end{tabular}       \\ \cmidrule(l){2-7} 
                                 & SchNOrb (950)                                                 & 50                                                 & 1478                                                 & 32                                                                       & $1 \times 10^{-4}$                                                                  & \begin{tabular}[c]{@{}l@{}}Decay LR by $\gamma=0.8$ after\\ 150 epochs w/o val loss improvement\end{tabular} \\ \midrule
\multirow{4}{*}{Uracil}          & Orbital Mixer (25K)                                                    & 50                                                 & 4500                                                 & 32                                                                       & $3 \times 10^{-4}$                                                                  & \begin{tabular}[c]{@{}l@{}}Decay LR by $\gamma=0.8$ every\\ $n_{\text{decay}}=1M$ steps\end{tabular}         \\ \cmidrule(l){2-7} 
                                 & SchNOrb (25K)                                                 & 500                                                & 4500                                                 & 48                                                                       & $1 \times 10^{-4}$                                                                  & \begin{tabular}[c]{@{}l@{}}Decay LR by $\gamma=0.8$ after\\ 15 epochs w/o val loss improvement\end{tabular}  \\ \cmidrule(l){2-7} 
                                 & Orbital Mixer (950)                                                    & 50                                                 & 4500                                                 & 32                                                                       & $3 \times 10^{-4}$                                                                  & \begin{tabular}[c]{@{}l@{}}Decay LR by $\gamma=0.8$ every\\ $n_{\text{decay}}=500K$ steps\end{tabular}       \\ \cmidrule(l){2-7} 
                                 & SchNOrb (950)                                                 & 50                                                 & 4500                                                 & 48                                                                       & $1 \times 10^{-4}$                                                                  & \begin{tabular}[c]{@{}l@{}}Decay LR by $\gamma=0.8$ after\\ 150 epochs w/o val loss improvement\end{tabular} \\ \bottomrule
\end{tabular}%
}
\caption{Training parameters and train/val/test splits used for training Orbital Mixer and the retrained SchNOrb models in this work.}
\label{tab:model_param}
\end{table*}

\section{Additional experimental results}

\begin{figure*}[h]
\centering
\includegraphics[width=\linewidth]{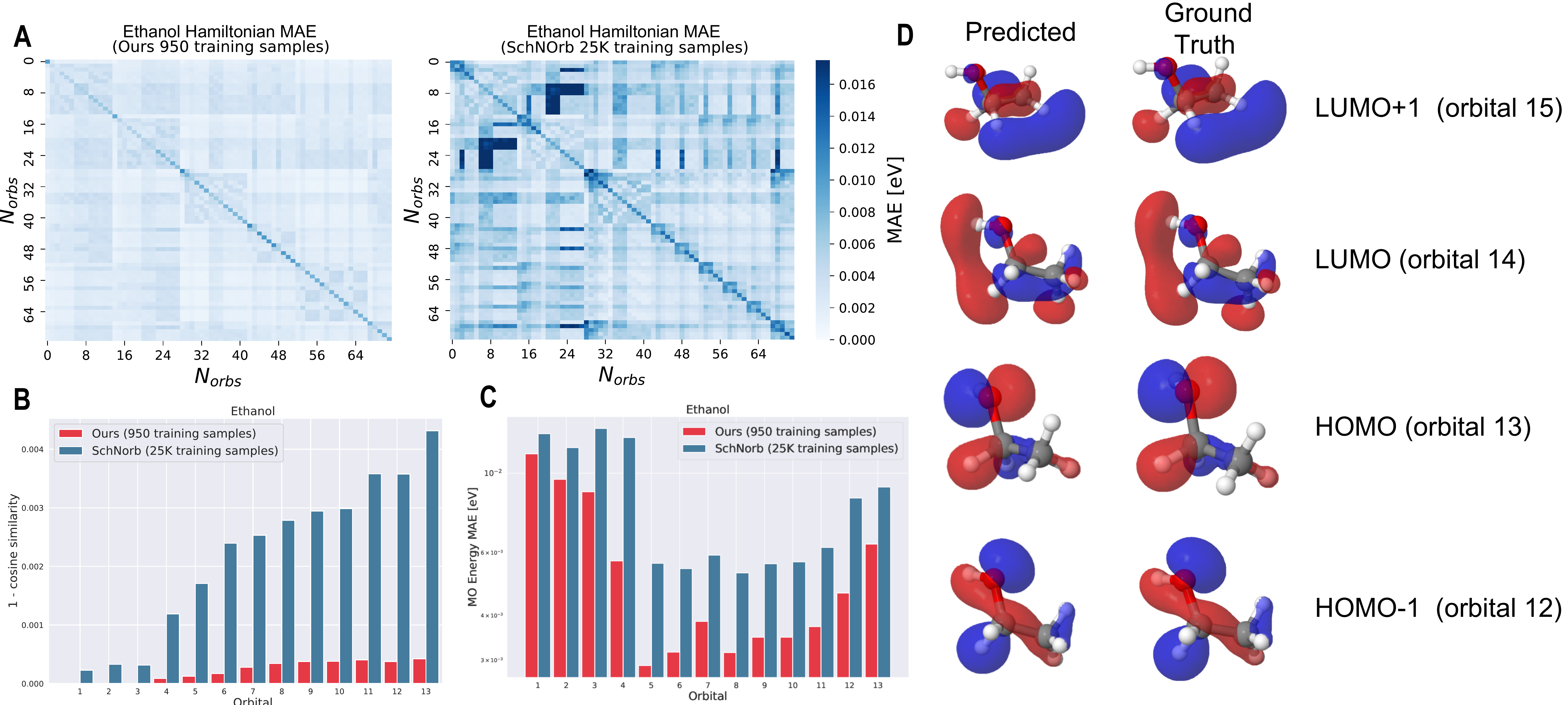}
\caption{Prediction errors for the Ethanol dataset of \textbf{(a)} the Hamiltonian $\mathbf{F}$, \textbf{(b)} the molecular orbital (MO) coefficients and \textbf{(c)} the MO energy between Orbital Mixer predictions trained with 950 configurations and SchNOrb trained with 25K configurations. \textbf{(d)} Visualization of Ethanol molecular orbital shapes derived from Orbital Mixer predicted and ground truth MO coefficients.}
\label{fig:ethanol_viz}
\end{figure*}

\begin{figure*}[h]
\centering
\includegraphics[width=\linewidth]{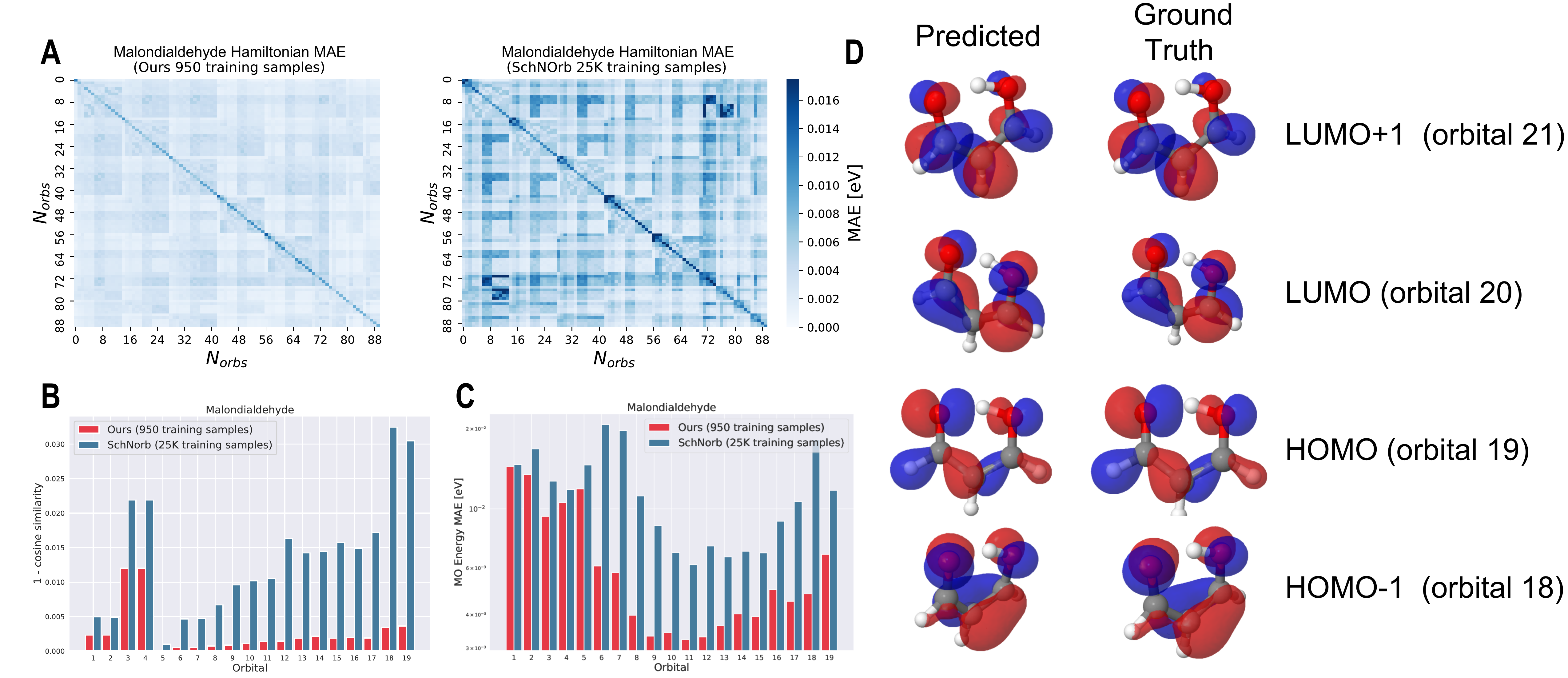}
\caption{Prediction errors for the Malondialdehyde dataset of \textbf{(a)} the Hamiltonian $\mathbf{F}$, \textbf{(b)} the molecular orbital (MO) coefficients and \textbf{(c)} the MO energy between Orbital Mixer predictions trained with 950 configurations and SchNOrb trained with 25K configurations. \textbf{(d)} Visualization of Malondialdehyde molecular orbital shapes derived from Orbital Mixer predicted and ground truth MO coefficients.}
\label{fig:mal_viz}
\end{figure*}

\begin{figure*}[htb!]
\centering
\includegraphics[width=\linewidth]{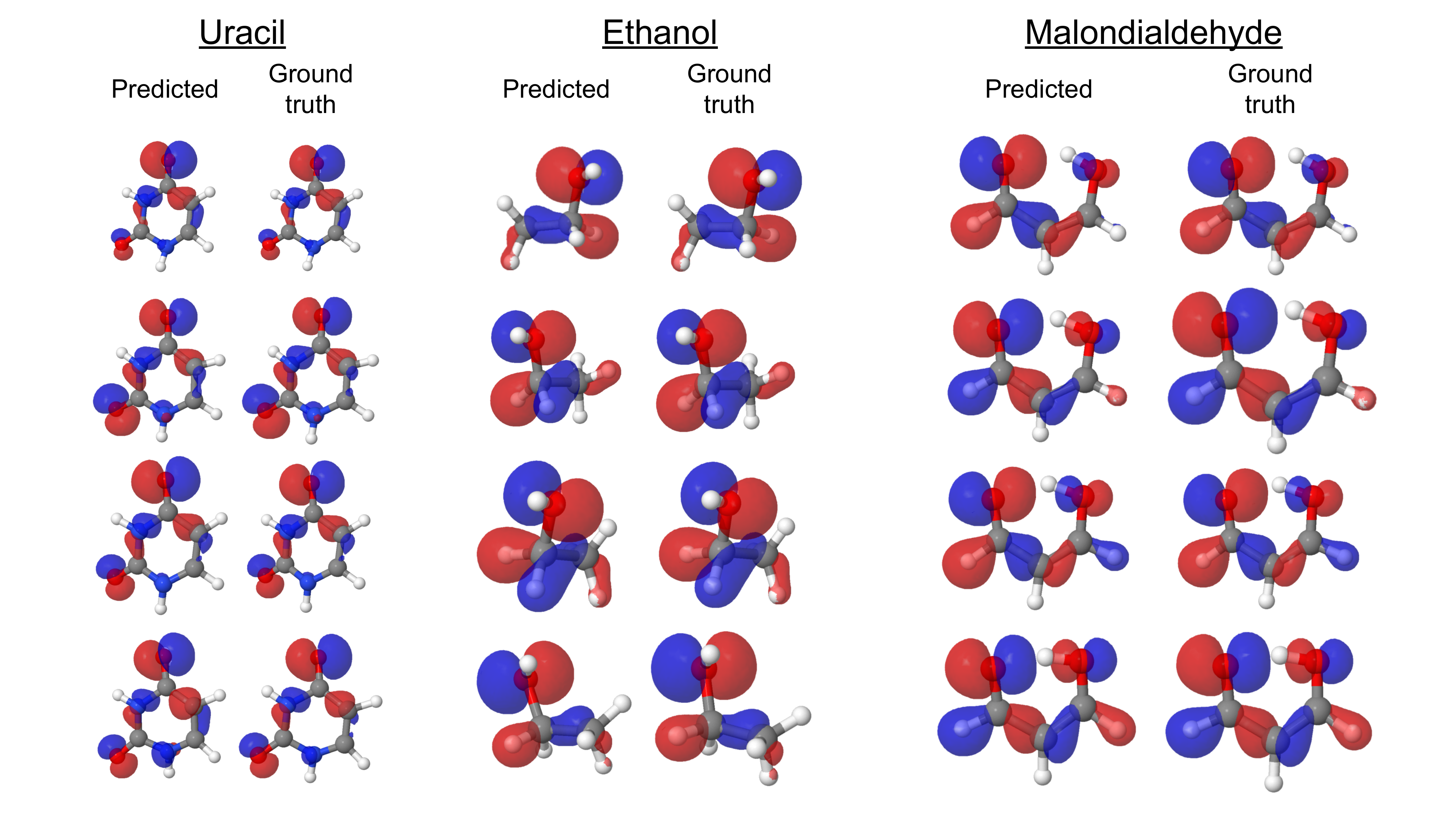}
\caption{Comparison of the highest occupied molecular orbital (HOMO) shapes for different molecular configurations derived from Orbital Mixer predicted and ground truth MO coefficients. Visualizations are shown for four random test configurations taken from the Uracil (left), Ethanol (center) and Malondialdehyde (right) datasets.}
\label{fig:orbstruct}
\end{figure*}

\begin{figure}[htb!]
\centering
\includegraphics[width=\linewidth]{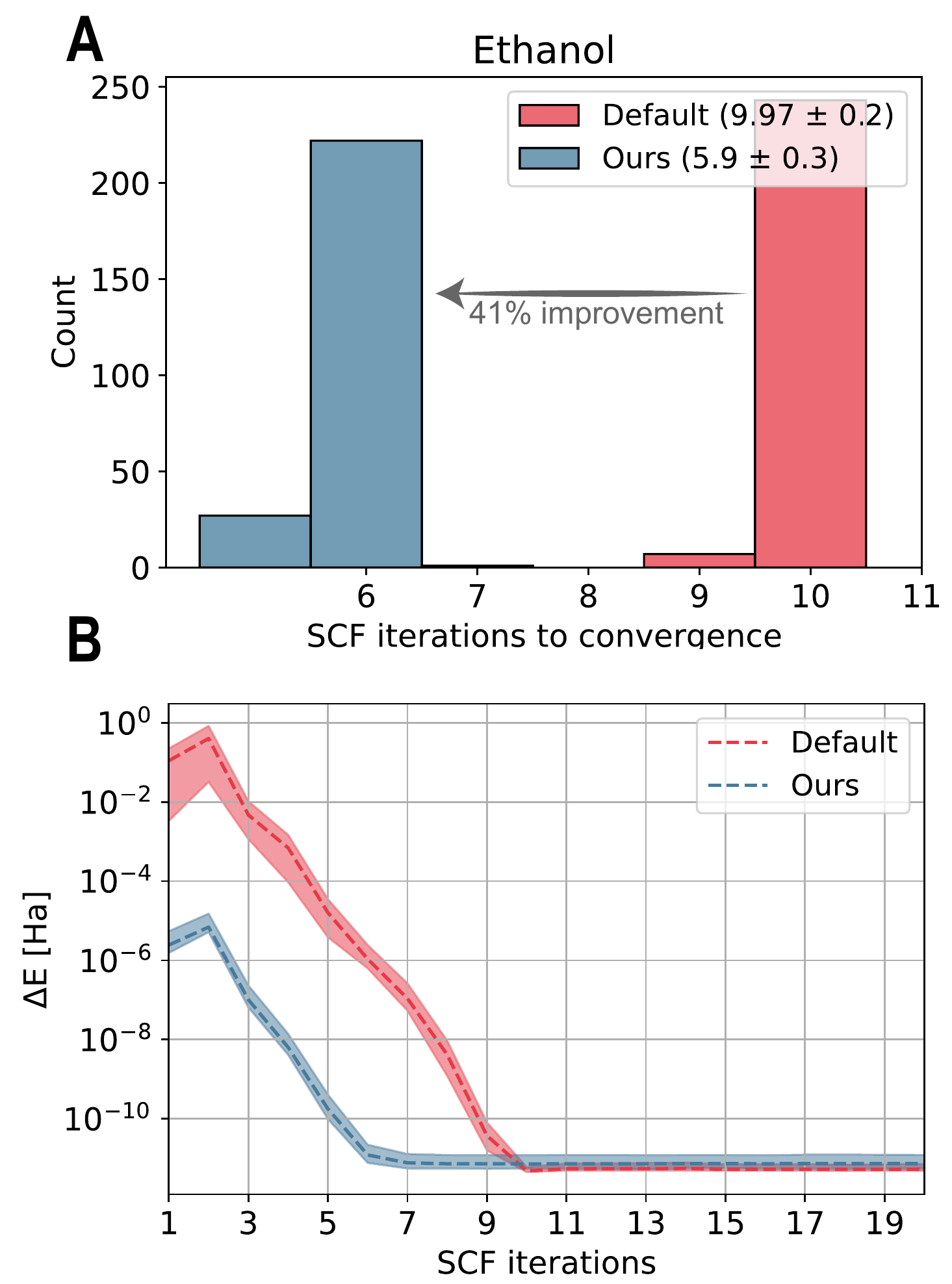}
\caption{\textbf{(a)} Number of SCF iterations required to reach convergence using initial guesses for the Hamiltonian $\mathbf{F}$ generated by Orbital Mixer and the default PySCF initialization for 250 test set Ethanol configurations. \textbf{(b)} Difference in the total energy at each SCF iterations compared to the terminal converged energy estimate. Errors about the mean dotted line bound the 25\% and 75\% quartiles.}
\label{fig:converge_ethanol}
\end{figure}

\begin{figure}[htb!]
\centering
\includegraphics[width=\linewidth]{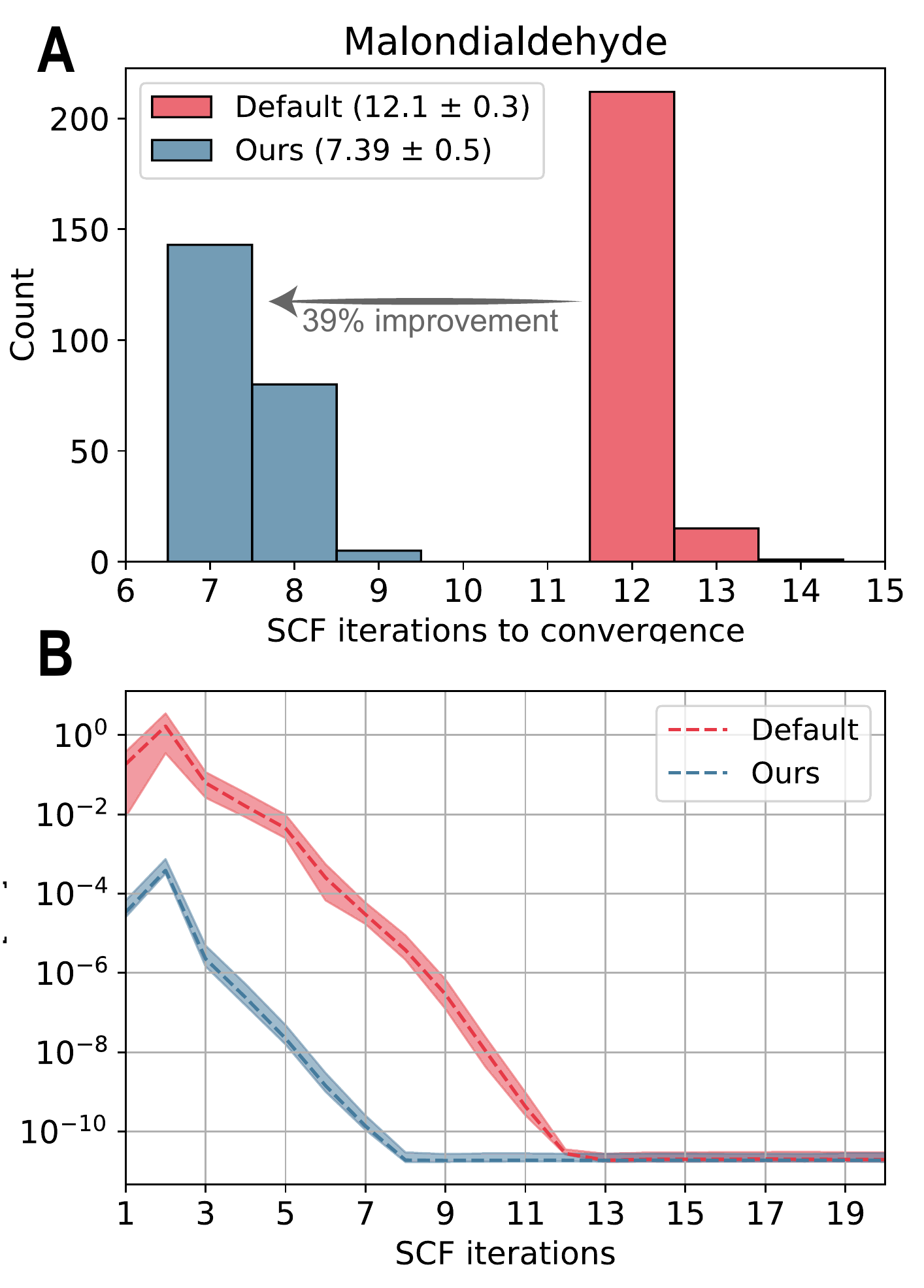}
\caption{\textbf{(a)} Number of SCF iterations required to reach convergence using initial guesses for the Hamiltonian $\mathbf{F}$ generated by Orbital Mixer and the default PySCF initialization for 228 test set Malondialdehyde configurations. \textbf{(b)} Difference in the total energy at each SCF iterations compared to the terminal converged energy estimate. Errors about the mean dotted line bound the 25\% and 75\% quartiles.}
\label{fig:converge_malondialdehyde}
\end{figure}

%\section{Do \emph{not} have an appendix here}
%
%\textbf{\emph{Do not put content after the references.}}
%%
%Put anything that you might normally include after the references in a separate
%supplementary file.
%
%We recommend that you build supplementary material in a separate document.
%If you must create one PDF and cut it up, please be careful to use a tool that
%doesn't alter the margins, and that doesn't aggressively rewrite the PDF file.
%pdftk usually works fine. 
%
%\textbf{Please do not use Apple's preview to cut off supplementary material.} In
%previous years it has altered margins, and created headaches at the camera-ready
%stage. 
%%%%%%%%%%%%%%%%%%%%%%%%%%%%%%%%%%%%%%%%%%%%%%%%%%%%%%%%%%%%%%%%%%%%%%%%%%%%%%%
%%%%%%%%%%%%%%%%%%%%%%%%%%%%%%%%%%%%%%%%%%%%%%%%%%%%%%%%%%%%%%%%%%%%%%%%%%%%%%%

\end{document}